# Spontaneous cross-section field in impurity graphene

M.B. Belonenko [1], N.G. Lebedev [2], A.V. Pak[2], N.N. Yanyushkina [2]

[1] Volgograd Institute of Business, Laboratory of nanotechnology,

400048 Volgograd, Russia

[2]Volgograd State University

400062 Volgograd, Russia

[2]e-mail: pak.anastasia@gmail.com, yana@inbox.ru

**Introduction**

Recently, graphene has attracted an interest to the nonlinear phenomena considerably stimulated creation of the materials, capable to show nonlinear properties in easily achievable experimentally conditions. One of such materials is graphene, representing the structure consisting of one layer of carbon atoms, and which has flat a hexagon lattice. The big electron mobility in graphene and its unique electrophysical characteristics draw to it attention, as one of alternatives of silicon base of modern microelectronics [1 - 4]. Note, that the electromagnetic waves extending in carbon structures become strongly nonlinear already at rather weak fields that involves possibility of distribution in carbon nanotubes and graphene electromagnetic solitary waves which are analogues solitons or even solitons. Discussed properties of the carbon structures have caused both the raised theoretical interest, and application attempts in devices of nonlinear optics [5]. Nonlinearity, according to the conclusions made in these works [6, 7], arises owing to change of classical distribution function of electrons and non-parabolic electron dispersion law.

A study of phase transitions represents one of the paradigms of modern fundamental physics. In particular, one of the major places in all variety of phase transitions occupies the non-equilibrium phase transition, owing to the occurrence to the application of different external fields.

It has theoretically been shown [8, 9], that in the presence of strong electric field the non-equilibrium phase transitions in an electronic gas in semiconductors with bulk-centered lattice are observed. The effect is shown in spontaneous occurrence of the cross-section field $E_y$ playing a role of an order parameter. Thus applied electric field $E_x$ directed along one of the symmetry axis of a crystal is command parameter. A necessary condition existence of a cross-section field, as shown in the given works, is non-additivity of an electronic spectrum: $\varepsilon(\boldsymbol{p}) \neq \varepsilon(p_x) + \varepsilon(p_y) + \varepsilon(p_z)$, where $\boldsymbol{p}$ – electron quasi-momentum (for example, an electronic spectrum in bulk-centered lattice becomes not additive $\varepsilon(p) \sim \cos(p_x a / 2\hbar) \cos(p_y a / 2\hbar) \cos(p_z a / 2\hbar)$ (вставь прибли-зительно равно) where $a$ – the lattice constant. Let's notice, that the condition non-additivity is carried out and in pure graphene, but physically more interesting graphene with impurity, as owing to that graphene [10] easily adsorbs to various atoms, and owing to that in impurity graphene there will be a raised electron concentration in a conductivity band that is important for practical applications. Besides, the spectrum should be limited, i.e. electron energy cannot be more than certain value.

All conditions set forth above are carried out for impurity graphene and it is possible to investigate possibility of a phase transitions existence in impurity graphene which is considered within the framework of the Anderson model that should be shown in occurrence cross-section components $E_x$ in the presence of field $E_y$ which will play a role of the command parameter.

Let us notice, that occurrence of a spontaneous cross-section field is important in the world of using of graphene as bases for transistors [11 - 13], notably: the arisen spontaneous cross-section field will reject electrons, that can essentially affect on current-voltage characteristic of the graphene transistor.

Summarising, one can draw a conclusion, that the research problem of the response of two-dimensional electronic system taking into account Anderson interaction of an impurity electrons and graphene electrons is represented enough actual.

### 1. The basic equations

Let us consider the response of graphene to external electric field. The problem geometry is given in fig. 1.

The Hamiltonian of the Anderson periodic models in a kind convenient for consideration of an electronic spectrum in graphene can be written as[14]:

$$H = \sum_{k\sigma} \varepsilon_k c_{k\sigma}^+ c_{k\sigma} + \sum_a \varepsilon_a \sum_\sigma a_\sigma^+ a_\sigma + U n_{a\sigma} n_{a-\sigma} + \sum_{ak\sigma} V_{ka} c_{k\sigma}^+ a_\sigma + \sum_{ak\sigma} V_{ka}^* a_\sigma^+ c_{k\sigma} \qquad (1)$$

Where $c_{k\sigma}$ $(c_{k\sigma}^+)$ and $a_\sigma$ $(a_\sigma^+)$ are the creation operator and annihilation operator of electrons in a crystal and atom accordingly, and $n_{a\sigma} = a_\sigma^+ a_\sigma$ is the operator of electron occupation number, $\varepsilon_a$ is the energy of electron, being on the adsorbed atom (adatom), $\varepsilon_k$ is the band energy of electrons in a crystal, $V_{ka}$ is the matrix element of electron transition from the adsorbed atom to graphene, $U$ is the coulomb energy of electron interaction in the adsorbed atom.

We calculate Hamiltonian parameters (1) on an example of atomic hydrogen which was considered as the impurity adsorbed on a graphene surface. The choice of the given type of an impurity is caused, in particular, by that in this case parameter $U \equiv 0$, as in atomic hydrogen only one electron.

The hybridization potential V = $V_{ka}$ in the Anderson model (1) was estimated from quantum-chemical approaches, as it is defined by overlap integral of wave functions of $s$-orbital (hydrogen atom) and $p_z$-orbital (carbon atom):

$$V = \frac{1}{2}\left(\beta_H + \beta_C\right) S_{HC},$$

$$S_{HC} = \int \Psi_{1s}(\mathbf{r}) \Psi_{2p_z}(\mathbf{r}) d\mathbf{r}, \qquad (2)$$

$$\Psi_{1s} = \frac{1}{\sqrt{\pi}}\left(\frac{z}{a_0}\right)^{\frac{3}{2}} e^{-\rho}, \rho = \frac{zr}{a_0}, z(H) = 1;$$

$$\Psi_{2p_z} = \frac{1}{4\sqrt{2\pi}}\left(\frac{z}{a_0}\right)^{\frac{3}{2}} \rho e^{-\frac{\rho}{2}} \cos\theta, \rho = \frac{zr}{a_0}, z(C) = 6;$$

Where $S_{HC}$ is the overlap integral of wave functions, $\beta_H$ and $\beta_C$ are the parameters calculated by semi empirical quantum -chemical method MNDO [15], $\beta_H$ =-6.99 eV, $\beta_C$ = - 7.93 eV, $a_0$ is the Bohr radius, $z$ is the atomic charge.

Estimations of hybridization potential give a value $V$ = - 1.43 эB. The energy is negative, therefore is formed a stable state.

For an estimation of the adsorbed atom $\varepsilon_d$ energy the method of the mirror images [16] is used. The following formula is resulted:

$$\tilde{\varepsilon}_a = I + \frac{1}{4\pi\varepsilon_0}\frac{e^2}{4l}$$

Where $I$ is the ionization potential of hydrogen atom, which equals -13.6 eV, $e$ is the elementary charge, $\varepsilon_0$ is the electric constant, $l$ = 1.2 Å is the distance between the adsorbed atom centre and a plane of its image in a substrate, which has been estimated from quantum -chemical semi empirical calculations.

After the estimation of Hamiltonian parameters (1) there is a problem about the description of a elementary excitation spectrum. We will take advantage for this purpose of a mathemat-ical apparatus of Green's functions [17] and a technique described in work [18]. So, for Fourier transform of one-partial Green's functions [18] of adsorbed atoms the motion equation is following:

$$E << a_\lambda \mid a_\lambda^+ >> = \frac{i}{2\pi} \left\langle \left[ a_\lambda, a_\lambda^+ \right]_+ \right\rangle + \sum_{k\sigma} \varepsilon_k << \left[ a_\lambda, c_{k\sigma} c_{k\sigma}^+ \right] \mid a_\lambda^+ >> +$$

$$+ \varepsilon_a \sum_\sigma << \left[ a_\lambda, a_\sigma a_\sigma^+ \right] \mid a_\lambda^+ >> + \sum_{k\sigma} V_{ka} << \left[ a_\lambda, c_{k\sigma}^+ a_\sigma \right] \mid a_\lambda^+ >> + \sum_{k\sigma} V_{ka}^* << \left[ a_\lambda, a_\sigma^+ c_{k\sigma} \right] \mid a_\lambda^+ >>$$

After calculations equation will be transformed to the following form ($\omega$ is the power variable):

$$\omega << a_\lambda \mid a_\lambda^+ >> = \frac{i}{2\pi} + \varepsilon_a << a_\lambda \mid a_\lambda^+ >> + \sum_k V_{ka}^* << c_{k\lambda} \mid a_\lambda^+ >>.$$

The equations on Green's function of a crystal look like:

$$E << c_{l\nu} \mid c_{l\nu}^+ >> = \frac{i}{2\pi} \left\langle \left[ c_{l\nu}, c_{l\nu}^+ \right]_+ \right\rangle + \sum_{k\sigma} \varepsilon_k << \left[ c_{l\nu}, c_{k\sigma} c_{k\sigma}^+ \right] \mid c_{l\nu}^+ >> +$$

$$\varepsilon_a \sum_\sigma << \left[ c_{l\nu}, a_\sigma a_\sigma^+ \right] \mid c_{l\nu}^+ >> + \sum_{k\sigma} V_{ka} << \left[ c_{l\nu}, c_{k\sigma}^+ a_\sigma \right] \mid c_{l\nu}^+ >> + \sum_{k\sigma} V_{ka}^* << \left[ c_{l\nu}, a_\sigma^+ c_{k\sigma} \right] \mid c_{l\nu}^+ >>$$

Similarly:

$$\omega << c_{l\nu} \mid c_{l\nu}^+ >> = \frac{i}{2\pi} + \varepsilon_l << c_{k\sigma} \mid c_{l\nu}^+ >> + V_{la} << a_\sigma \mid c_{l\nu}^+ >>$$

In the result expression for Green's function of a crystal lattice taking into account the adsorbed atom of defect will be written as:

$$<< c_{k\sigma} \mid c_{k\sigma}^+ >> = \frac{i}{2\pi} \frac{(\omega - \varepsilon_a)}{(\omega - \varepsilon_a)(\omega - \varepsilon_k) - |V_{ka}|^2} \qquad (5)$$

Analytical expression for Green's function for the graphene crystal lattice (5) allows us define eigenvalues of electrons energy in a crystal, caused by adsorption of atomic hydrogen. The eigenvalues of electrons energy a crystal lattice with the attached defects atoms give poles of Green's functions [18]:

$$E(k) = \frac{1}{2} \left[ \varepsilon_a + \varepsilon_k \pm \sqrt{(\varepsilon_a - \varepsilon_k)^2 + 4 \frac{N_{imp}}{N} |V_{ka}|^2} \right] \qquad (6)$$

Where $\varepsilon_k$ is the band structure of graphene, $N$ is the number of unit cells, $N_{imp}$ is the number of impurity atoms.

Let us consider the dispersion law describes graphene properties, and enters in (6), looks like [13]:

$$\varepsilon(k) = \gamma \sqrt{1 + 4\cos(ap_x)\cos\left(\frac{ap_y}{\sqrt{3}}\right) + 4\cos^2\left(\frac{ap_y}{\sqrt{3}}\right)} \qquad (7)$$

Where $\gamma \approx 2.7\,\text{eV}, a = \dfrac{3b}{2\hbar}$, $b = 0.142$ nm is the distance between the neighbouring carbon atoms in the graphene, $k = (p_x, p_y)$.

Further, we take advantage of the «average electron» method and we will consider, that the motion equation can be written down in the following form [14]:

$$\frac{d\overline{p}}{dt} = q\overline{E} \qquad (8)$$

Where $q$ is the electron charge.

With calibration using: $\vec{E} = -\dfrac{1}{c}\dfrac{\partial \vec{A}}{\partial t}$ and for a case of zero temperatures, it is possible to receive:

$$p_x = p_{0x} + qE_x t$$
$$p_y = p_{0y} + qE_y t$$

It should be noted:

$$\upsilon(p_x) = \frac{\partial E(p_x, p_y)}{\partial p_x}, \quad \upsilon(p_y) = \frac{\partial E(p_x, p_y)}{\partial p_y}$$

Further, according to the «average electron» method we use a current definition [15]:

$$j = \int_0^\infty q\upsilon(\overline{p}(t))\exp(-t)dt, \qquad (9)$$

Where $j$ is the current density, $q$ is the electron charge, $\overline{p}(t)$ is the decision of the equation (8), and relaxation time is equal unity.

For impurity graphene it is convenient to present a dispersion law in the series form:

$$E(p_x, p_y) = \sum_{m,n} A_{mn} \cos(mp_x) \cos(np_y),$$

$$A_{mn} = \frac{1}{(2\pi)^2} \int_{-\pi}^{\pi} \int_{-\pi}^{\pi} \varepsilon(p_x, p_y) \cos(mp_x) \cos(np_y) dp_x dp_y$$

In this case we receive the expression for $x$ - components of the current density:

$$j_x = \sum_{m,n} A_{mn} \left( \cos\left(\frac{2\pi m}{3}\right) + \cos\left(\frac{\pi k}{3}\right) \cdot (-1)^i \right) \frac{m^2 E_x (n^2 E_y^2 - 1 - m^2 E_x^2)}{\left(1 + (nE_y + mE_x)^2\right)\left(1 + (nE_y - mE_x)^2\right)} \quad (10)$$

The transverse field $E_x$ is defined by boundary conditions for the given applied field $E_y$. Let us assume that a circuit is opened in the x-direction:

$$j_x = 0 \quad (11)$$

This condition corresponds to some solve for the transverse field: $E_x = E_x(E_y)$. The equation (11) has two decisions:

$$E_x = 0,$$

$$\sum_{m,n} A_{mn} \left( \cos\left(\frac{2\pi m}{3}\right) + \cos\left(\frac{\pi k}{3}\right) \cdot (-1)^i \right) \frac{m^2 \left(n^2 E_y^2 - 1 - m^2 E_x^2\right)}{\left(1 + (nE_y + mE_x)^2\right)\left(1 + (nE_y - mE_x)^2\right)} = 0 \quad (12)$$

The transverse field spontaneously appears in one of two mutually opposite directions at the some values of parameters in the second equation of (12). In this case, we deal with non-equilibrium one-order phase transition. The appearance of the transverse field component represents perhaps the simplest example of self-organization in graphene.

## 2. Results of the numerical analysis

Typical dependence $j_x$ on value $E_x$ is described (10) is represented in fig. 2.

From the resulted dependence it is visible, that besides a usual site with negative differential conductivity which is peculiar to all substances with the periodic dispersion law, on current-voltage characteristic there is a site with absolute negative conductivity that it is possible to connect with non-equilibrium electrons systems in the impurity graphene, caused first of all, strong non-parabolic of the dispersion law. We note, that the similar condition can lead to splitting carbon nanotubes on the domains which detailed consideration is beyond this work.

Dependence cross-section components of field $E_x$ on $E_y$ which is defined as the nonzero decision of the equation (12), is shown on fig. 3, 4.

It worth to note, that similar dependence on parameter $\varepsilon_a$ is weakly shown up, and more important that there are two nonzero decisions. One of these decisions (smaller on the module) is thermodynamically unstable.

The constant field not arises from zero values, so the arising effect in terms of the phase transitions theory, should be interpreted as first order phase transition. Occurrence of a constant field can be connected with a site on a current-voltage characteristic with an absolute negative conductivity, because electrons system is the non-equilibrium redistribution of electrons in momentum space is occured hereby that an arising field to aspire to compensate action of a cross-section field.

The transverse field $E_x$, which is emergent spontaneously, can be thermodinamically unstanble, as opposite to always stable solve for open circuit in the x-direction $E_x$ =0.

We also investigate the stability using the method proposed in [6]. We introduct the following function:

$$\Phi(E_x) = \int_{0}^{E_x} j_x(E_x')dE_x' + \text{constant} \qquad E_y = \textit{fixed} \qquad (13)$$

The given function is usually called with synergetics potential and it is analogue of thermodynamic potential for nonequilibrium problems. According to [6] stability conditions of the decision is:

$$\frac{d\Phi}{dE_x} = 0; \qquad \frac{d^2\Phi}{dE_x^2} > 0; \qquad (14)$$

This actually means, that in the given nonequilibrium situation function (13) reaches the minimum in a stationary condition, and thus, function $\Phi$ can act as analogue of thermodynamic potential for equilibrium systems. So dependence of "potential" $\Phi$ on field $E_x$, for a number of values $E_y$ is displayed on fig. 5.

It can be seen that the function $\Phi$ has two minimum poinst and two maximum points. It should be noted, maximum corresponds to a smaller in module solve of equation (12), and minimum corresponds to a larger in module solve. We will remark, that dot branches in fig. 3. and in fig. 4. correspond to the maximum of function $\Phi$ (the unstable decision), and continuous and dotted branches correspond to the minimum of function $\Phi$ (the steady decision).

This transition, in which the electric field appears spontaneously, is concerned to ferroelectric type. Though, the transverse field $E_x$ plays the role of the order parameter, and the field $E_y$ is the analogue of temperature (controlling parameter).

### 3. The conclusion

In summary we will formulate our main results:

1. The occurrence of the electric field perpendicular to the applied external electric field in impurity graphene, described within the framework of the Anderson model is revealed.

2. Most strongly spontaneously arising electric field depends on the value of a matrix element of electron transition from the adsorbed atom in graphene.

3. The analysis of synergetic potential has shown that the arising condition with a spontaneous cross-section field is steady.

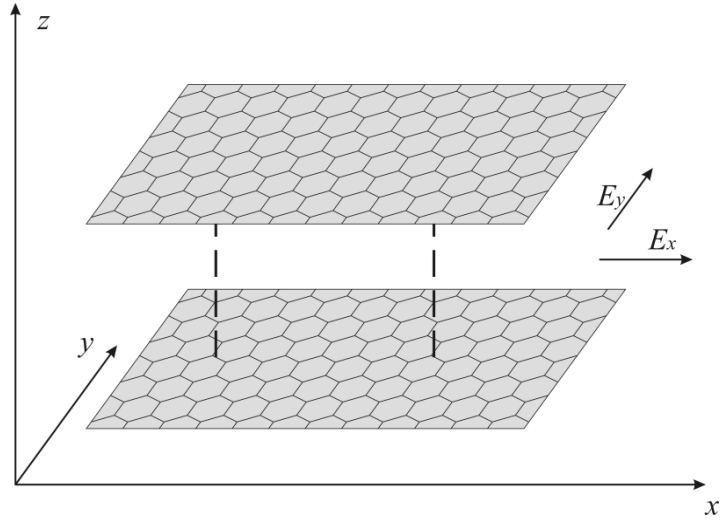

Fig. 1. Geometry of the problem.

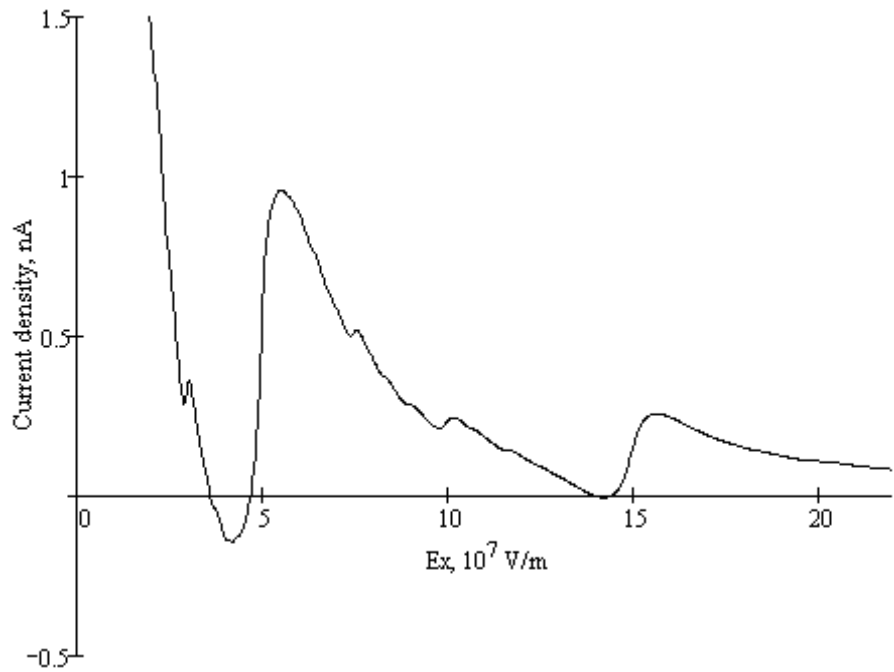

Fig.2. Dependence of current density on the field $E_x$, the field $E_y$ is fixed ($E_y$ =5·10$^7$ V/m).

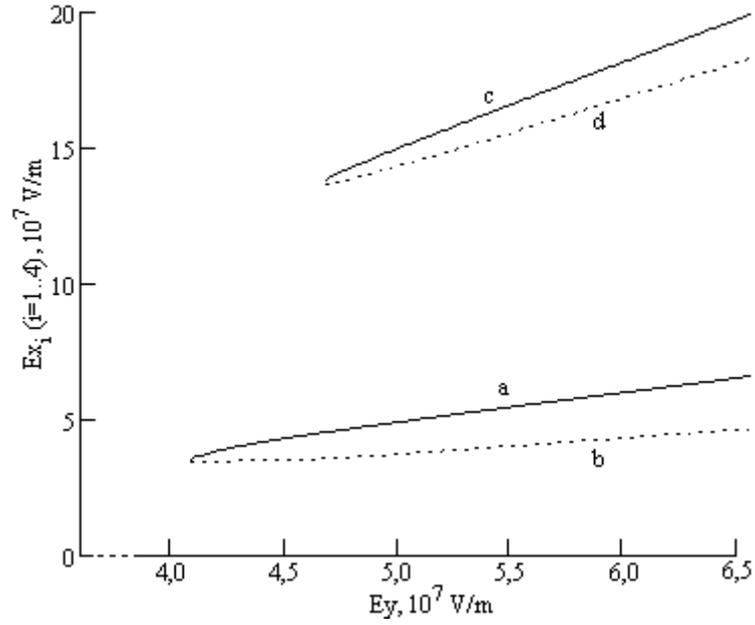

Fig.3. Dependence of the field $E_x$ on field $E_y$: $\varepsilon_a$=1.43 eV, V=10.0 eV. Branches (a) and (c) correspond to a stable solution, and branches (b) and (d) correspond to an unstable solution.

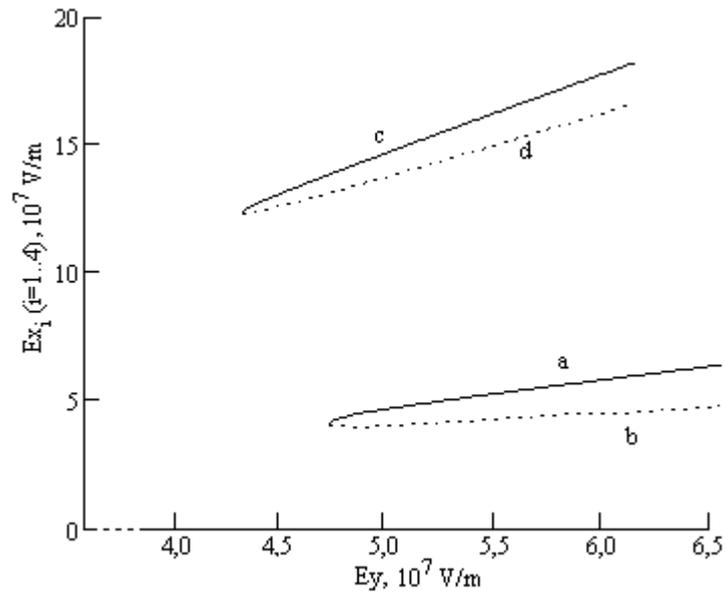

Fig.4. Dependence of the field $E_x$ on field $E_y$: $\varepsilon_a$=1.43 eV, V=5.0 eV. Branches (a) and (c) correspond to stable solution, and branches (b) and (d) correspond to an unstable solution.

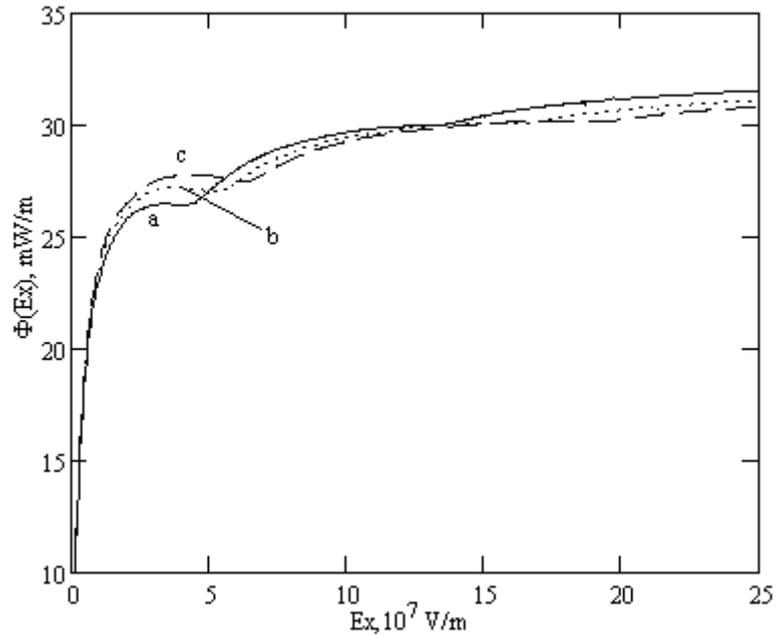

Fig.5. Dependence of the function $\Phi$ on field $E_x$ at a fixed value of field $E_y$: (a) $E_y$ =4.5·10$^7$ V/m; (b) $E_y$ =5.5·10$^7$ V/m; (c) $E_y$ =6.5·10$^7$ V/m.